\documentclass[12pt]{article}

\usepackage{graphics}
\usepackage{amssymb}
\usepackage{cite}
\usepackage{epsfig}
\usepackage{psfrag}

\def\eq#1{(\ref{#1})}
\def\Eq#1{Eq.~(\ref{#1})}
\newcommand{\secn}[1]{Section~\ref{#1}}

\def\beq{\begin{equation}}
\def\eeq{\end{equation}}
\def\beqa{\begin{eqnarray}}
\def\eeqa{\end{eqnarray}}
\def\bet{\begin{tabular}}
\def\eet{\end{tabular}}
\def\del{\partial}
\newcommand{\ex}[1]{{\rm e}^{#1}} \def\ii{{\rm i}}
\newcommand{\Tr}{{\rm Tr}}
\newcommand{\sect}[1]{\setcounter{equation}{0}\section{#1}}

\renewcommand{\a}{\alpha}

\newcommand{\e}{\epsilon}
\newcommand{\ve}{{\vec{\e}}}

\newcommand{\bxi}{\mbox{\boldmath $\xi$}}
\newcommand{\bkp}{{\bf k}_\perp}
\newcommand{\bxp}{{\bf x}_\perp}
\newcommand{\byp}{{\bf y}_\perp}
\renewcommand{\Im}{{\rm Im}\,}

\begin{document}

\begin{titlepage}

\setcounter{page}{0}

\begin{flushright}
{CERN-PH-TH/2004-244}\\
{DFTT-04-29}\\
{QMUL-PH-04-14}
\end{flushright}

\vspace{0.6cm}

\begin{center}
{\Large \bf Two-loop Euler-Heisenberg effective actions from charged 
open strings} \\

\vskip 0.8cm

{\bf Lorenzo Magnea}\\
{\sl Dipartimento di Fisica Teorica, Universit\`a di Torino and}\\
{\sl INFN, Sezione di Torino, Via P. Giuria 1, I-10125 Torino, Italy,}\\
{\sl and CERN, Department of Physics, TH division}\\
{\sl CH--1211 Geneva 23, Switzerland}\\

\vskip .3cm

{\bf Rodolfo Russo\footnote{On leave of absence
from {\it Queen Mary, University of London}, E1 4NS London, UK}}\\
{\sl CERN, Department of Physics, TH division}\\
{\sl CH--1211 Geneva 23, Switzerland}\\

\vskip .3cm

{\bf Stefano Sciuto}\\
{\sl Dipartimento di Fisica Teorica, Universit\`a di Torino}\\
 and {\sl  INFN, Sezione di Torino}\\
{\sl Via P. Giuria 1, I-10125 Torino, Italy}\\

\vskip 1.2cm

\end{center}

\begin{abstract}

We present the multiloop partition function of open bosonic
string theory in the presence of a constant gauge field strength, and
discuss its low-energy limit. The result is written in terms of
twisted determinants and differentials on higher-genus Riemann
surfaces, for which we provide an explicit representation in the
Schottky parametrization. In the field theory limit, we recover from
the string formula the two-loop Euler-Heisenberg effective action for
adjoint scalars minimally coupled to the background gauge field.

\end{abstract}

\vfill

\end{titlepage}

\sect{Introduction}
\label{intro}

The dynamics of charged particles in a constant electromagnetic
background has been a focus of considerable theoretical interest since
the early days of quantum field theory. To recall just two important
contributions, Euler and Heisenberg~\cite{Heisenberg:1935qt} computed
the one-loop QED effective action by integrating out fermion fields,
and later Schwinger~\cite{Schwinger:1951nm} derived the probability of
pair creation in a constant electric field, focusing on the absorptive
part of the one-loop calculation (see~\cite{Dunne:2004nc} for a recent
review and a detailed list of references). A similar problem can be
studied in string theory, and also in this context it has provided
several important insights. Open strings, in particular, represent a
natural generalization of charged particles since they couple, through
their endpoints, to a gauge field. Bachas and
Porrati~\cite{Bachas:1992bh} generalized Schwinger's computation to
the case of open (super)strings, and later the same technique was
applied to study the $T$-dual case of moving
D-branes~\cite{Bachas:1995kx} and to a finite temperature 
environment~\cite{Tseytlin:1998kw}.

In this paper we will focus on the case of open bosonic strings and we
will study the multiloop partition function in the presence of a
constant Yang-Mills field strength $F$. We will then use the string
formula in the low-energy limit to recover the Euler-Heisenberg
effective action at one and two loops, considering specifically the
coupling of the gauge field to adjoint scalars. Even for the bosonic
string, the explicit formulation of this partition function requires
some new input, beyond the known results of multiloop perturbative
string theory. From the mathematical point of view, the open string
diagram is represented as usual by a Riemann surface with $g + 1$
boundaries, however the presence of an external field $F$ introduces
twisted boundary conditions along some of the boundaries. As a
consequence, the basic geometric building blocks of the string
amplitude, such as the determinant of the Laplace operator on the
Riemann surface, are deformed by $F$. The necessary ingredients to
derive the multiloop partition function for charged open strings were
assembled in~\cite{Russo:2003tt,Russo:2003yk}, developing earlier
studies~\cite{Verlinde:1986kw,Alvarez-Gaume:1987vm}. We note also that
twisted boundary conditions in the open string channel correspond to
cuts along homology cycles for closed strings. This suggests that the
present formalism might have broader applications: the appearence of
cuts on the Riemann surface representing a string amplitude, in fact,
is a generic feature of the Ramond-Neuveu-Schwarz formalism. The
Ramond sector of the superstring, for instance, has a square-root cut
for fermion fields, while closed strings on orbifolds have $n$-fold
cuts in the twisted sector. Thus it is not surprising to see the same
mathematical objects appearing in different contexts (compare, for
instance,~\cite{Aoki:2003sy,Russo:2003yk,Antoniadis:2004qn}).

Even if the mathematical formulation of the string effective action is
written in the language of two-dimensional Riemannian geometry, its
physical content is very close to the quantum field theory
counterpart. In fact it is expected, and we will explicitly show, that
the two results should be precisely connected in the field theory
limit, when the typical length of the open string $\sqrt{\a'}$ is sent
to zero.

The idea to construct a precise mapping between the low energy
behaviour of string theory and the corresponding field theory Feynman
diagrams is very old~\cite{Scherk:1971xy}, and was applied to
effective actions in~\cite{Metsaev:1987ju}; it received renewed
interest when it was noted~\cite{Mangano:1987xk} that the organization
of tree-level gluon amplitudes suggested by string theory is an
efficient computational tool for high energy processes in
QCD. Subsequently, the methods to perform the field theory limit of
string amplitudes at loop level were pioneered by Bern and
Kosower~\cite{Bern:1991aq}, who studied one-loop amplitudes in gauge
theories. String-inspired techniques were then used to obtain novel
results, relevant to collider
phenomenology~\cite{Bern:1993mq,Bern:1996je}. Later, the
correspondence between string diagrams and ordinary Feynman diagrams
was made more precise, and it was shown that one-loop Yang-Mills
amplitudes could be recovered from open bosonic string theory on a
diagram by diagram basis~\cite{DiVecchia:1996uq}, and even completely
off-shell~\cite{Frizzo:2000ez}.  While at one loop these techniques
have been very successful for Yang-Mills theory, and have also been
applied to gravity~\cite{Bern:1993wt,Dunbar:1994bn}, extension to two
loops for gauge bosons has proven
difficult~\cite{Magnea:1997kv,Kors:2000bb}. It has however been shown
that one can tune the field theory limit of bosonic strings to
reproduce Feynman diagrams in scalar theories, at one and two
loops~\cite{DiVecchia:1996kf}, and with both cubic and quartic
vertices~\cite{Frizzo:1999zx,Marotta:1999re}.

In the second part of this paper we will make use some of these
techniques to study the low energy regime of the charged string
partition function at one and two loops. We will take the simplest
field theory limit~\cite{Frizzo:1999zx,Marotta:1999re}, isolating the
contribution of the string ground state (the tachyon). As expected,
the Euler-Heisenberg effective action will arise from the string
formula in the scaling limit in which the dimensionful physical gauge
field is kept constant as $\a'\to 0$.

\begin{figure}
\begin{center} 
\leavevmode
\hbox{%
\epsfxsize=13cm 
\psfrag{tatau}{$\tau \to - \tau^{-1}$}
\epsfbox{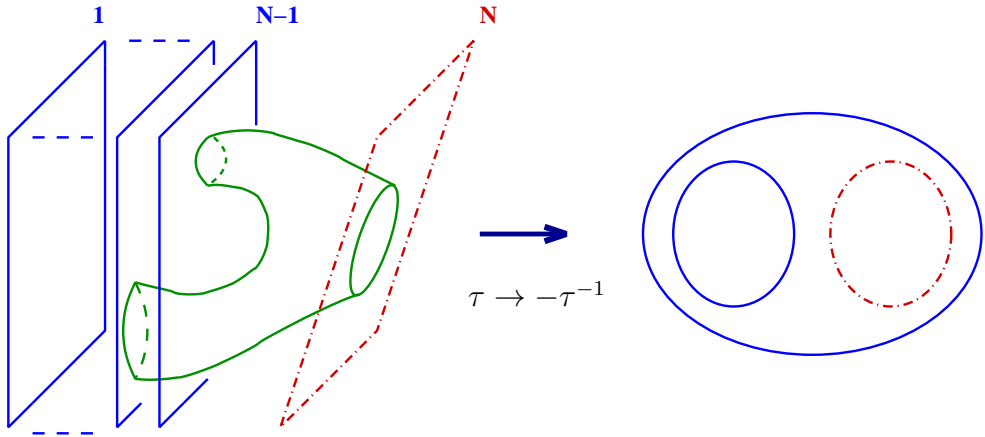}}
\end{center} 
\caption{A graphical representation of the modular transformation from
a configuration of closed strings attached to a charged $D$-brane
(picked in a stack of $N$) to open strings with a charged boundary.}
\label{branes}
\end{figure}

The structure of the paper is the following. In \secn{bospart} we
begin by recalling the computation of vacuum diagrams for ordinary
bosonic strings, and then move on to study the modifications due to
the presence of a constant gauge field strength. The crucial point is
that when $F$ is constant the world-sheet equations of motion of the
string are unchanged and only the boundary conditions are modified. A
powerful method to take into account these twisted boundary conditions
is the boundary state formalism, which describes {\em closed} strings
propagating between the various boundaries. To get our final result,
we must then perform the modular transformation exchanging $a$ and $b$
homology cycles, leading to an expression for the amplitude in the
open string channel, as described pictorially in Fig.~\ref{branes}. In
order to be self-contained and to keep track of all normalizations, we
give in \secn{field} the derivation of the two-loop Euler-Heisenberg
effective action in the appropriate field theory.  Finally, in
\secn{low} we perform the field theory limit described
in~\cite{Frizzo:1999zx,Marotta:1999re} on the $g = 1,2$ string
partition function, and we show that it precisly reproduces the
results of \secn{field}. The final Section discusses further
applications of our results and some possible developments.

\sect{Higher-genus effective actions for bosonic strings}
\label{bospart}

\subsection{Partition functions without external fields}
\label{vacen}

The partition function of closed bosonic string theory at $2$ and $3$
loops was first derived in~\cite{Belavin:1986tv,Moore:1986rh}, using
previous mathematical results on modular forms. This derivation,
however, is not directly generalizable beyond genus $g = 3$. In fact,
only for $g \leq 3$ one can choose to parametrize the moduli space of
inequivalent Riemann surfaces by using the elements of the period
matrix $\tau_{\mu\nu}$, with $\mu\leq \nu$, as independent
parameters.  An alternative derivation of the bosonic partition
function was presented in~\cite{DiVecchia:1987uf}, where the operator
approach was used. The final expression is automatically written in
the Schottky parametrization of the Riemann surface, so that it is
valid for any $g$, in a framework which however has the drawback of
blurring the modular properties of the results.

As usual, one can derive the open string partition function from the
closed string result by requiring that the period matrix be compatible
with the involution~\cite{Blau:1987pn,Bianchi:1988fr} that squeezes a
closed genus-$g$ surface into an open one. In our case we can obtain
the disk with $g$ holes by restricting the period matrix to be purely
imaginary. In the Schottky parametrization the genus-$g$ partition
function for the open bosonic string, with Neumann boundary
conditions, is given by
\beqa
Z (g) & = & \int \frac{1}{d V_{a b c}} \prod_{\mu = 1}^g \left[
d k_\mu d \eta_\mu d \xi_\mu 
\frac{(1 - k_\mu)^2}{k_\mu^2 (\xi_\mu - 
\eta_\mu)^2} \right] 
\left[\det \left(\Im \tau \right) \right]^{-\frac{d}{2}} \nonumber \\
&& \times \, {\prod_\alpha}' \left[\frac{
\prod_{n = 2}^\infty (1 - k_\alpha^n)^2}
{\prod_{n = 1}^\infty (1 - k_\alpha^n)^{d} }\right]~.
\label{scg}
\eeqa
Here $d$ is the space-time dimension, $d = 26$ for the ordinary
bosonic string. Moduli space is parametrized by the multipliers
$k_\mu$ and by the fixed points $\xi_\mu$ and $\eta_\mu$ of the $g$
projective transformations forming the basis for the Schottky group at
genus $g$. We refer to the Appendices of~\cite{Russo:2003tt} for a
short explanation of the Schottky parametrization and for all the
conventions and notations we use in this regard. Here we only note
that the primed product in \eq{scg} is over primary classes of the
Schottky group, characterized by elements with multipliers $k_\a$.
The factor $d V_{abc}$ accounts for the volume of the projective group
which leaves the measure invariant, corresponding to the freedom to
fix arbitrarily three fixed points. At two loops, we will take
advantage of this freedom by setting $\eta_1 = 0$, $\xi_1 = \infty$
and $\xi_2 = 1$.

\Eq{scg} can be compared, for $g = 2$, with the results of 
of~\cite{Belavin:1986tv,Moore:1986rh},
\beq
Z (2) = \int d \tau_{11} d \tau_{12} d \tau_{22}
\left[\det \left(\Im \tau \right) \right]^{- d/2} 
\frac{1}{\pi^{12} \chi_{10}(\tau)}~,
\label{bm}
\eeq
where $\chi_{10}$ is the unique modular form of weight~$10$ with no
zeros, and is equal to the product of the squares of the ten even
$\theta$ functions at genus $g = 2$, $\chi_{10} = \prod_{m \, {\rm even}}
\theta_m^2(\tau)$. Note that the three moduli appearing in \eq{scg}
for $g = 2$ ($k_1$, $k_2$ and $\eta \equiv \eta_2$) are related to the
elements of the period matrix. In particular, in the limit where the
Riemann surface degenerates into a graph ($k_i \to 0$), we have
\beq
2\pi \ii \, \tau_{11} = \log{k_1}  + O(k)~~,~~~
2\pi \ii \, \tau_{22} = \log{k_2}  + O(k)~~,~~~
2\pi \ii \, \tau_{12} = \log{\eta}  + O(k)~.
\label{tautok}
\eeq
The two expressions~\eq{scg} and~\eq{bm} were shown to agree
perturbatively (when expanded for small $k_i$)
in~\cite{Petersen:1988ux,Roland:1988hg}. In order to see that they are
exactly equivalent, one can check that both of them follow from the
same `first principles' formula~\cite{D'Hoker:1988ta},
\beq
Z(g) = \int \prod_{a = 1}^{3 g - 3} d m_a \, \frac{\det
\langle \mu_j^{(a)} | \phi_k \rangle} {\sqrt{\det
\langle \phi_j | \phi_k \rangle}} \, {\rm det}' (\partial^\dagger
\partial) \, \left[\det \left(\Im \tau \right) \right]^{- d/2} 
Z_1^{- d}~,
\label{fp}
\eeq
which can be derived directly from Polyakov path integral. Here
$\mu_j^{(a)}$ is a system of Beltrami differentials related to the
moduli $m_a$; $\partial$ is the operator appearing in the ghost
Lagrangian and acting on a $(b,c)$ system of weight $(2,-1)$; $\phi_j$
is a basis of periodic and holomorphic differentials of weight $2$;
finally, $Z_1$ is related to the partition function of a single chiral
boson, for which one can find explicit expression in Eq.~(7.3)
of~\cite{D'Hoker:2001qp}, or one can use the Schottky parametrization,
where $Z_1 = {\prod_\alpha}' \prod_{n = 1}^\infty (1 -
k_\alpha^n)$. To prove the equivalence of \eq{scg} and \eq{bm}, the
basic idea is to make two different choices for the moduli $m_a$, and
thus for the bases of Beltrami and quadratic differentials in~\eq{fp},
and show that they give rise to~\eq{bm} and~\eq{scg} respectively. The
two results must then be equal, since~\eq{fp} does not depend on the
particular choice made for the moduli or for the differentials. A few
steps of this proof are summarized in Appendix A.

So far, we have described the partition function by using the open
string point of view, where the world-sheet looks like a disk with $g$
holes.  In this case the moduli $\{\tau_{\mu \nu}\}$ (or $\{k_i,
\eta\}$) are directly related to the lengths of the various strips
representing the open strings propagating in the string diagram. An
alternative approach is to adopt a closed string description for the
same amplitude. In this picture the world-sheet has the shape of a
disk glued to $g$ cilinders, whose boundaries are described by
boundary states (for a review
see~\cite{DiVecchia:1999rh,DiVecchia:1999fx}). In this channel the
partition functions reads~\cite{Frau:1997mq}
\beq
Z^c (g) = \int \frac{1}{d V_{abc}} \prod_{\mu = 1}^{g} \left[
\frac{d q_\mu \, d^2 \eta^c_\mu \, (1 - q_\mu)^2}{q_\mu^2 \, 
(\eta^c_\mu - \bar\eta^c_\mu)^2} \right]
{\prod_{\a}}' \left(\prod_{n = 1}^\infty (1 - q^n_\a)^{-d}
\prod_{n = 2}^\infty (1 - q^n_\a)^2 \right) \, ,
\label{mis}
\eeq
where $d V_{abc}$ again signals that we have to fix three real
variables among the $\eta$'s. The superscript $c$ is a reminder of the
fact that the parameters appearing in~\eq{mis} are appropriate for
describing closed string exchanges among the various
boundaries. Notice the absence of factors of $\det \left(\Im \tau
\right)$ in this formulation.

At the level of the Schottky parametrization the modular
transformation connecting the closed string~\eq{mis} and the open
string~\eq{scg} channels is rather complicated. In fact, the relation
between the multipliers $q$ and $k$ is non-analytic, since $2 \pi \ii
\tau_{11}^c \sim \log{q_1}$ and $2 \pi \ii \tau_{11} \sim \log{k_1}$,
while the open and closed string period matrices are connected by
means of the usual modular transformation $\tau^c = - \tau^{-1}$. In
order to transform~\eq{mis} into~\eq{scg}, one must first rewrite the
integrand in terms of geometrical objects with simple modular
properties, such as $\theta$-functions. Then it is possible to perform
the modular transformation by using the known transformation
properties of these functions, as done
in~\cite{Russo:2003tt,Russo:2003yk}.  On the other hand, of course,
the modular transformation $\tau^c = - \tau^{-1}$ can be directly
performed on~\Eq{bm}: the result is again the same expression, now
written as a function of $\tau^c$, but again without any factor of
$\det \left(\Im \tau \right)$.

\subsection{Open strings in a constant background field}
\label{backem}

The results reviewed in the previous section are appropriate for open
strings with Neumann boundary conditions along all boundaries.  We
will now outline the derivation of the partition function for {\it
charged} open strings, {\it i.e.} open strings with mixed boundary
conditions
\beq
\left[ \del_\sigma X^i + \ii \,\del_\tau X^j 
F_j^{~i \,(A)} \right]_{\sigma = 0} = 0~,
\label{gbc}
\eeq
where $F$ is a constant gauge field strength and $A = 0, \ldots, g$
labels the boundary on which the boundary condition~\eq{gbc} is
imposed ($A = 0$ being the external boundary). $F$ can always be put
in a block-diagonal form, so for the sake of simplicity we will take
the space-time indices to be in the plane $i,j = 1,2$.  For charged
strings at least one of the differences $F^{(\mu)} - F^{(0)}$, for
$\mu = 1, \ldots, g$, is nonvanishing.

A direct computation of the charged partition function in the open
string channel is difficult, mainly because the string coordinates
have a non-integer mode expansion. It is possible to sew with a
propagator two charged states of the $3$-string vertex, but the
result~\cite{Chu:2002nd} is rather complicated and it is difficult to
proceed and build multiloop diagrams. Here we will recall the
derivation presented in~\cite{Russo:2003yk}, where an alternative
approach was followed. The idea is to compute the string diagram in
the closed string channel by using boundary states satisfying
\beqa
\left( \partial_\tau X^1 + {\rm i} \, 
\tan(\pi \epsilon^A) \; \partial_\sigma X^2
\right)_{\tau = 0} | B \rangle_{F_A} & = & 0~,
\nonumber \\
\left( \partial_\tau X^2 - {\rm i} \, 
\tan(\pi \epsilon^A) \; \partial_\sigma X^1
\right)_{\tau = 0} | B \rangle_{F_A} & = & 0~.
\label{bcond}
\eeqa
These boundary conditions are just a rewriting of those in~\Eq{gbc}
after the exchange $\tau \leftrightarrow - \sigma$ and with the
convention $F_{12}^{(A)} = - F_{21}^{(A)} = \tan(\pi
\epsilon^A)$. With the change of coordinates $X^\pm =
\frac{1}{\sqrt{2}} (X^1 \pm \ii X^2)$, the constraints~\eq{bcond}
become diagonal and the computation of vacuum diagrams is almost
identical to that of~\cite{Frau:1997mq}. The only difference is that
the matrices ${\cal S}^A$ appearing there\footnote{These matrices are
denoted by $S$ in Ref.~\cite{Frau:1997mq}; here we label them ${\cal
S}$ to distinguish them from the Schottky group generators introduced
below.} in the boundary state contain, in the plane $\{X^1,X^2\}$,
some $\epsilon$-dependent phases, instead of having all elements equal
to $\pm 1$. To be specific, we take the external boundary to have
Neumann boundary conditions ($F^{(0)} = 0$), so that $\vec{\e}$ is a
vector with $g$ components, denoted by $\e_\mu$, encoding the values
of the gauge field on the remaining $g$ boundaries. The explicit form
of the matrices\ ${\cal S}_{\mu}$ appearing in the boundary state is
then ${\cal S}_\mu = {\rm diag} \{ {\rm e}^{2 \pi \ii \epsilon_\mu},
{\rm e}^{- 2 \pi \ii \epsilon_\mu}\}$. It is not difficult to follow
these phases through the computation. One verifies that their effect
is simply to modify the contribution of the oscillators in the charged
plane $\{X^1,X^2\}$ to the partition function. The result is of the
form
\beq
Z_F^c (g) = \left( \prod_{\mu = 1}^g \frac{1}{\cos \pi \e_\mu} \right) 
\int \left[ d Z \right]_g^c \, {\cal R}_g \left(q_\a, \vec{\e}\right)~,
\label{clch}
\eeq
where $\left[d Z \right]_g^c$ is the integrand in~\Eq{mis},
representing the $F = 0$ result, while the $\vec{\e}$ dependence is
encoded in the factor
\beq
{\cal R}_g \left(q_\a, \vec{\e}\right) =  \frac{{\prod_\alpha}' 
\prod_{n = 1}^\infty (1 - q_\alpha^n)^2}{ {\prod_\alpha}' 
\prod_{n = 1}^\infty \left( 1 - \ex{ - 2 \pi \ii \vec{\e} \cdot \vec{N}_\a} 
q^n_\alpha \right) \left( 1 - \ex{ 2 \pi \ii \vec{\e} \cdot \vec{N}_\a} 
q^n_\alpha \right)}~.
\label{ratioclosed}
\eeq
Here $\vec{N}_\a$ is a vector with $g$ integer entries: the $\mu^{\rm
th}$ entry counts how many times the Schottky generator $S_\mu$ enters
in the element of the Schottky group $T_\a$, whose multiplier is
$q_\a$ (for example $S_\mu$ contributes~$1$, while $(S_\mu)^{-1}$
contributes~$-1$). The factors of $1/\cos(\pi \e)$ in \Eq{clch},
finally, are nothing but a rewriting of the Born-Infeld contribution
to the boundary state normalization (see for
instance~\cite{DiVecchia:1999fx}). For $g = 1$ \Eq{clch} agrees with
the results of~\cite{Tseytlin:1986zz,Tseytlin:1986ti}, as one can see
by using $\zeta$-function regularization to rewrite $1/\cos(\pi \e)$ as 
$\prod_{n = 1}^\infty (1 + F^2)^{-1}$.

Now we would like to perform the modular transformation $\tau^c = -
\tau^{-1}$ on~\eq{clch}, in order to write the effective action in the
presence of a nonvanishing $F$ in the open string channel. We already
know from the previous section that $\left[d Z \right]_g^c$ transforms
into the integrand of~\Eq{scg}, which we denote by $\left[d Z
  \right]_g$. Thus we only need to transform the factor ${\cal R}_g
\left(q_\a, \vec{\e}\right)$, which contains the dependence on the
external field $F$, and to write it in terms of the open strings
variables $k_\mu$, $\eta_\mu$ and $\xi_\mu$. To do this, we follow the
same approach discussed in the previous section. First, we rewrite the
products over the Schottky group in terms of geometrical objects with
simple transformation properties under the modular group, like
$\theta$ functions, differentials and the prime form; then, we perform
the modular transformation; as a last step, we go back to the Schottky
parametrization, which is the most appropriate for performing the low
energy limit. The technical tool needed in this derivation is the
higher-genus generalization of the Jacobi formulae expressing $\theta$
functions as products. These formulae can be derived by exploiting
bosonization identities in two dimensions, as done
in~\cite{Pezzella:1988jr,Losev:1989fe,DiVecchia:1989ht}. Details are
given in~\cite{Russo:2003tt,Russo:2003yk}, where the presence of the
twists $\vec{\e}$ is also taken into account.  The results of
Refs.~\cite{Russo:2003tt,Russo:2003yk} can be written as
\beq
{\cal R}_g \left(q_\a, \vec{\e} \right) = \left(\ex{ 2 \pi \ii \e_g} - 1 
\right) {\cal R}_g \left(k_\a, \vec{\e} \cdot \tau \right) \;
\ex{- \ii \pi \vec{\e} \cdot \tau \cdot \vec{\e}} \;
\frac{\det \left(\tau \right)}{\det \left(\tau_\ve \right)}~,
\label{dqdk}
\eeq
where we have assumed $\e_g \neq 0$.  The only new object appearing
in~\Eq{dqdk} is the matrix $\tau_\ve$. As suggested by the notation, it
is an $\ve$-dependent generalization of the period matrix. Recall that
the matrix elements of $\tau$ are the periods along the $b$ cycles of
the normalized Abelian differentials
\beq
\frac{1}{2 \pi \ii} \int_{b_\nu} \omega_\mu = \tau_{\nu \mu}~, \qquad
\frac{1}{2 \pi \ii} \int_{a_\nu} \omega_\mu = \delta_{\nu \mu}~.
\label{defper}
\eeq
Similarly, $\tau_\ve$ can be expressed in terms of the periods of {\em
twisted} meromorphic differentials (known as Prym differentials). In
our case the twists are along the $a$ cycles and are fixed by
$\ve$, so they depend on the external gauge field. An explicit
expression for the matrix $\tau_\ve$ was derived
in~\cite{Russo:2003tt,Russo:2003yk}. Begin by defining $g$
$\ve$-dependent differentials, as
\beqa
\zeta_\mu^{\vec{\e}} (z) & = & {\sum_\a}^{(\mu)} 
\ex{ 2 \pi \ii (\vec{\e} \cdot \vec{N}_\a
+ \e_\mu)} \left[\frac{1}{z - T_\a (\eta_\mu)} -
\frac{1}{z - T_\a(\xi_\mu)} \right]
\nonumber \\ & + &
(1 - \ex{ 2 \pi \ii \e_\mu}) \sum_\a \ex{ 2 \pi \ii \vec{\e} 
\cdot \vec{N}_\a} \left[\frac{1}{z - T_\a(z_0)} - \frac{1}{z - 
T_\a(a_\mu^\a)} \right]~.
\label{zeta}
\eeqa 
The first sum runs over all elements of the Schottky group, except
those whose rightmost generator is $S_\mu^{\pm 1}$, while the second
sum is unrestricted; furthermore, in the second line, $a_\mu^\a =
\eta_\mu$ if $T_\a$ is of the form $T_\a = T_\beta S_\mu^l$ with $l
\geq 1$, while $a_\mu^\a = \xi_\mu$ otherwise. These differentials are
characterized by the following features: first, they are twisted along
the $b$ cycles, {\it i.e.} they obey $\zeta_\mu^{\vec{\e}} \left(
S_\nu (z) \right) d S_\nu (z) = \exp \left(2 \pi \ii \e_\nu \right)
\zeta_\mu^{\vec{\e}} (z) d z$; next, they are holomorphic everywhere
except at the arbitrarily chosen point $z = z_0$, where they have a
single pole with residue $(1 - \ex{2 \pi \ii \e_\mu})$; finally, they
reduce, in the $\ve \to 0$ limit, to the usual Abelian differentials
normalized as in~\eq{defper}. There are $g$ independent differentials
satisfying these requirements in agreement with Riemann-Roch
theorem. The matrix $\tau_\ve$ is defined as the usual period matrix,
with the abelian differentials substituted by Prym differentials,
where however the twist is placed along the $a$
cycles~\cite{Russo:2003yk}, in order to take into account the modular
transformation to the open string configuration. Explicitly,
\beq
(\tau_\ve)_{\nu \mu} = \frac{1}{2 \pi \ii} \int\limits_{w}^{S_\nu (w)}
d z \left[ \zeta_\mu^{\vec{\e} \cdot \tau} (z) \ex{\frac{2 \pi \ii}{g - 1} 
\vec{\e} \cdot \vec{\Delta}_z } \right]~,
~~ (\nu \not= g)~; \qquad
\left(\tau_\ve \right)_{g \mu} = \ex{2 \pi \ii (\vec{\e} 
\cdot \tau)_\mu} - 1~,
\label{taue}
\eeq
where $\vec{\Delta}_z$ is the vector of Riemann constants (or Riemann class)
whose definition in the Schottky parametrization is
\beq
\Delta_z^\mu = \frac{1}{2 \pi \ii} \left[ - \frac{1}{2} \log k_\mu + 
\ii \pi + \sum_{\nu = 1}^g  \hspace{-5mm}{\phantom{\sum_\a}}^{(\nu)} 
{\sum_\a \,}' \hspace{-5mm}{\phantom{\sum_\a}}^{(\mu)} \log 
\left( \frac{\xi_\nu - T_\a(\eta_\mu)}{\xi_\nu - 
T_\a(\xi_\mu)} \, \frac{z - T_\a (\xi_\mu)}{z - T_\a (\eta_\mu)} 
\right)\right]~.
\label{ricla}
\eeq
Here the sum over $T_\a$ excludes those with a power of $S_\nu$ to 
the left and those with a power of $S_\mu$ to the right; moreover, the
identity is excluded for $\mu = \nu$.

A few comments are now in order. First of all, notice that the matrix
elements of $\tau_\ve$ actually do not depend on the base point $w$,
as one can easily check by taking a derivative of \eq{taue} with
respect to $w$ and using the periodicity of the integrand along the
$b$ cycles. Next, we remark that the asymmetry of the last line of
$\tau_\ve$ is just apparent. The integrals of the twisted
differentials along the $b$ cycles, in fact, are not independent: a
linear combination of them is fixed by the value of the residue at the
pole, as one can see by integrating them along the cycle $\prod_\mu
a_\mu b_\mu a_\mu^{-1} b_\mu^{-1}$. Using this linear dependence one
gets
\beq
(\tau_\ve)_{g \mu} = \ex{-\frac{2 \pi \ii}{g - 1} 
\vec{\e} \cdot \vec{\Delta}_{z_0} } \sum_{\nu = 1}^g
\frac{\ex{2 \pi \ii \e_\nu} - 1}{2 \pi \ii} \int_{b_\nu} d z
\left[ \zeta_\mu^{\vec{\e} \cdot \tau} (z)
\ex{\frac{2 \pi \ii}{g - 1} \vec{\e} \cdot \vec{\Delta}_z }\right]~.
\label{sc1}
\eeq
As a consequence, if desired, one could replace \Eq{dqdk} with the 
more symmetric expression
\beq
{\cal R}_g \left(q_\a, \vec{\e} \right) = {\cal R}_g \left(k_\a, 
\vec{\e} \cdot \tau \right) \;
\ex{- \ii \pi \vec{\e} \cdot \left(\tau \cdot \vec{\e} - \frac{2}{g - 1}
\vec{\Delta}_{z_0} \right)}\;
\frac{\det \left(\tau \right)}{\det \left(\hat{\tau}_\ve \right)}~,
\label{dqdk2}
\eeq
where
\beq
(\hat{\tau}_\ve)_{\nu \mu} = \frac{1}{2 \pi \ii} \int\limits_{w}^{S_\nu (w)}
d z \left[ \zeta_\mu^{\vec{\e} \cdot \tau} (z) \ex{\frac{2 \pi \ii}{g - 1} 
\vec{\e} \cdot \vec{\Delta}_z } \right]~, \qquad \mu,\nu = 1, \ldots, g~. 
\label{symtaue}
\eeq
From the computational point of view, the formulation of~\Eq{taue} is
somewhat easier to use, and we will adopt it in what follows.  The
final observation is that the dependence of $\tau_\ve$ on the position
of the pole, $z_0$, disappears when one takes the determinant. In fact,
as discussed in~\cite{Russo:2003tt}, it is possible to rewrite ${\rm
det}(\tau_\ve)$ as the determinant of a $(g-1) \times (g-1)$ matrix
containing linear combinations of $\zeta_\mu^{\vec{\e}}$ in which the
pole has cancelled. These linear combinations can then be interpreted
as holomorphic Prym differentials with periodicities fixed by
$\vec{\e}$. As the Riemann-Roch theorem dictates, there are only $g -
1$ holomorphic Prym differentials. Using the formulation of \Eq{symtaue},
on the other hand, one finds that $\det (\hat{\tau}_\ve)$ has a dependence 
on $z_0$ which cancels the explicit dependence arising in \Eq{dqdk2}
through the Riemann class $\vec{\Delta}_{z_0}$.

We are now in a position to write a completly explicit expression for
the string effective action in the open string channel. It is
\beq
Z_F (g) =  \left(\frac{\ex{ 2 \pi \ii \e_g} - 1}{\prod_{\mu = 1}^g 
\cos{\pi \e_\mu}} \right) 
\int \left[d Z \right]_g \, \left[ \ex{- \ii \pi \vec{\e} \cdot \tau \cdot
\vec{\e}} \; \frac{\det \left(\tau \right)}{\det \left(\tau_\ve \right)}
\; {\cal R}_g \left(k_\a, \vec{\e} \cdot \tau \right) \right]\,.
\label{effstr}
\eeq

This expression can be easily generalized to the case where the gauge
field strength is non-trivial also in other space-time directions. It
is sufficient to introduce a different $\vec{\e}$ for each charged
plane, and add in~\eq{effstr} other $\vec{\e}$-dependent factors
exactly like those in parenthesis. If the field is of electric type,
one has to use an imaginary $\vec{\e}$, or simply substitute $\vec{\e}
\to \ii \vec{\e}$ in all the formulae presented in this section.
\Eq{effstr} will be our starting point in taking the low-energy limit
in \secn{low}.

\sect{Euler-Heisenberg effective action for scalar fields}
\label{field}

\subsection{Adjoint scalars in a constant background field}
\label{conve}

To make direct contact with the string formalism, let us consider a
$U(N)$ gauge field ${\cal A}_\mu$, represented as a hermitean $N
\times N$ matrix ${\cal A}_\mu = \sum_a A_\mu^a \, T_a$, where $T_a$
($a = 0, \ldots, N^2 - 1$) are $U(N)$ generators satisfying
\beq
[T_a, T_b] = {\rm i} f_{a b c} \, T^c~; \qquad \Tr \left( T_a T_b \right) =
\frac{1}{2} \, \delta_{a b}~.
\label{gennorm}
\eeq
We will treat ${\cal A}_\mu$ as a classical background gauge field
coupled to a quantum massive scalar fields, also in the adjoint
representation, $\Phi = \sum_a \varphi^a \, T_a$.  The covariant
derivative then acts on the matrix $\Phi$ as
\beq
D_\mu \Phi = \partial_\mu \Phi + {\rm i} [\Phi, {\cal A}_\mu]~,
\label{covder}
\eeq
where we absorbed a factor of the gauge coupling in the normalization
of the field ${\cal A}_\mu$. The lagrangian we will consider (with the 
normalizations of~\cite{Frizzo:1999zx}) is then
\beq
{\cal L} = \Tr \left[ D_\mu \Phi D^\mu \Phi - m^2 \Phi^2 + \frac{2}{3}
\lambda \, \Phi^3 \right]~.
\label{lag1}
\eeq
The gauge field configuration corresponding to a single charged brane
in the string picture is a diagonal ${\cal A}_\mu$ matrix, with all
eigenvalues vanishing except one, which we take to be the last one,
$\left({\cal A}_\mu \right)_{A B} = A_\mu \, \delta_{A,N}
\delta_{B,N}$. In order to have a constant field strength ${\cal
F}_{\mu \nu}$ corresponding to a constant chromomagnetic field in the
$x_3$ direction, we then pick $A_\mu = B \, x_1 \, g_{\mu 2}$.  This
choice of background breaks the symmetry in color space, so that the
matter ``multiplet'' will have both neutral and charged components
with respect to ${\cal A}_\mu$. One can write
\beq
\Phi (x) = \frac{1}{\sqrt{2}} \left( \matrix{ \sqrt{2} \, \Pi (x)  & 
\bxi (x) \cr \bxi^\dagger (x) &  \sigma(x) } \right)~, 
\label{splitphi}
\eeq
where $\Pi$ is a hermitean $(N - 1) \times (N - 1)$ matrix
representing a field in the adjoint representation of $U(N - 1)$,
$\bxi$ is complex vector in the fundamental representation of $U(N -
1)$, while $\sigma$ is a singlet real field. With the normalizations
given by \eq{splitphi}, all fields are canonically normalized and the
lagrangian \eq{lag1} becomes
\beqa
{\cal L} & = & \Tr \left[ \partial_\mu  \Pi \, \partial^\mu \Pi 
\right] + \frac{1}{2} \partial_\mu \sigma \, \partial^\mu \sigma +
D_\mu \bxi^\dagger D^\mu \bxi - m^2 \, \Tr \left( \Pi^2 \right) - \frac{1}{2}
m^2 \sigma^2 \nonumber \\ & - & m^2 \bxi^\dagger \bxi + \frac{2}{3} \lambda
\, \Tr \left( \Pi^3 \right) + \sqrt{2} \, \frac{\lambda}{6} \sigma^3 + 
\frac{\lambda}{\sqrt{2}} \, \sigma \bxi^\dagger \bxi + \lambda
\, \bxi^\dagger \Pi \, \bxi~.
\label{lag2}
\eeqa
Clearly, $\sigma$ and $\Pi$ are neutral with respect to ${\cal
A}_\mu$, whereas $\bxi$ and $\bxi^\dagger$ are oppositely charged,
with an abelian covariant derivative defined by
\beq
D_\mu \bxi = \partial_\mu \bxi + {\rm i} A_\mu \bxi~.
\label{coveder}
\eeq
The presence of the external field affects only the propagation of the
charged field $\bxi$, so the effective action in the background of
${\cal A}_\mu$ can be determined with the ordinary Feynman rules, upon
replacing the free $\bxi$ propagator with the one computed in the
chosen background. The form of the propagator of a scalar field in a
constant electromagnetic background has been known for a long
time~\cite{Fock:1937dy,Schwinger:1951nm}, and we give a derivation
suitable for our purposes in Appendix B. The result for the
coordinate-space propagator, expressed as a Schwinger parameter
integral, is given by
\beqa
& & \hspace{-15pt} G_\xi (x, y) \, = \, \frac{1}{(4 \pi)^{d/2}} \, 
{\rm e}^{- \frac{{\rm i}}{2} B (x_1 + y_1) (x_2 - y_2)} \int_0^\infty d t 
\, {\rm e}^{- t m^2} \, t^{- d/2 + 1} \, 
\frac{B}{\sinh (B t)} \times \label{fincoo} \\ & & \hspace{-18pt}
\exp \left[ 
\frac{1}{4 \, t} \left( \left( x_0 - y_0 \right)^2 - \left(\bxp - \byp 
\right)^2 \right) - \frac{B}{4 \tanh(B t)} \left( \left( x_1 - 
y_1 \right)^2 - \left(x_2  - y_2 \right)^2 \right) \right] ,
\nonumber
\eeqa
where we denote by $\bxp$ the $(d - 3)$-dimensional position vector
orthogonal to the $(x_1,x_2)$ plane. \Eq{fincoo} can be used directly
to construct vacuum diagrams, as described below.

\subsection{Vacuum diagrams}
\label{vacdia}

Vacuum diagrams contributing to the effective action for the field
$\bxi$ can be computed in an intuitively appealing way using the
coordinate-space representation of the charged propagator,
\Eq{fincoo}. At one loop, for example, the only contributing diagram
is a $\bxi$ loop without interaction vertices. Recalling that at one
loop the vacuum diagram is not directly related to the effective
action, but rather to its derivative with respect to the
mass, one can simply write
\beq
W_\xi^{(1)} (m, B) \, = \, - \int d m^2 \int d^d x \, G_\xi 
\left( x, x \right)~, 
\label{onedef}
\eeq
where the overall factor of $1/2$ related the square root arising from
the semi-classical Gaussian integration has been cancelled by the fact
that we are dealing with a complex field, involving two real degrees
of freedom. Using \Eq{fincoo}, one immediately obtains the well known
result~\cite{Weisskopf:1996bu}
\beq
W_\xi^{(1)} (m, B) \, = V_d \, \frac{N - 1}{(4 \pi)^{d/2}} \int_0^\infty
d t \, {\rm e}^{- t \, m^2} \, t^{- d/2} \, \frac{B}{\sinh(B t)}~,
\label{oneres}
\eeq
where $V_d$ is the volume of space-time and the factor $N - 1$ simply
counts the components of $\bxi$ circulating in the loop. Notice that
the integration with respect to the mass squared in~\eq{onedef} simply
adds a factor of $1/t$ to the powers contained in the
propagator~\eq{fincoo}. Notice also that here and below we will be
concerned only with bare diagrams, and we will not discuss the
inclusion of renormalization counterterms.

\begin{figure}
\begin{center} 
\leavevmode
\hbox{%
\epsfxsize=9cm
\psfrag{xi}{$\bxi$}
\psfrag{pi}{$\Pi$}
\psfrag{si}{$\sigma$}
\epsfbox{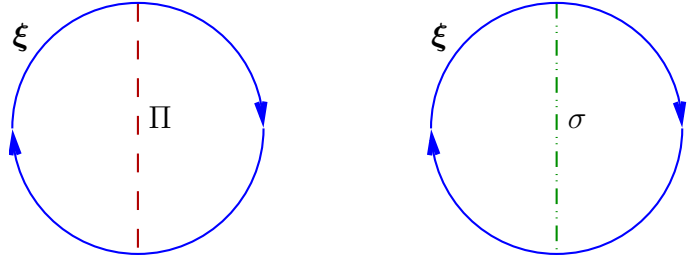}}
\end{center}
\caption{Two-loop irreducible diagrams with charged propagators.}
\label{irre}
\end{figure}

At two loops, the effective action is given directly by the sum of
one-particle-irreducible vacuum diagrams. The nontrivial contribution
in this case comes from diagrams involving a $\bxi$ loop, shown in
Fig.~(\ref{irre}). To compute these diagrams, we need the Feynman
rules for the $\bxi-\Pi$ and the $\bxi-\sigma$ vertices, derived from
the lagrangian in \Eq{lag2}. Parametrizing the matrix field $\Pi$ as
$\Pi = \sum_a \pi_a t^a$, with $t_a$ the $U(N - 1)$ generators in the
fundamental representation, the $\bxi-\Pi$ vertex is simply given by
${\rm i} \lambda (t^a)_{B C}$. Similarly, the $\bxi-\sigma$ vertex is
given by ${\rm i} \lambda \delta_{B C}$. Taking into account our
normalization of group generators, and a symmetry factor equal to
$1/2$, one finds that the first diagram in Fig.~(\ref{irre})
contributes
\beq
W_{\xi \Pi}^{(2)} (m, B) \, = \, - \, \lambda^2 \frac{(N - 1)^2}{4} 
\int d^d x \, d^d y \, G_\xi \left( x, y \right) \, 
G_\xi \left( y, x \right) \, G_\Pi 
\left( x, y \right)~,
\label{twodef}
\eeq
where $G_\Pi$ is the ordinary free scalar propagator. A short
calculation gives
\beq
W_{\xi \Pi}^{(2)} (m, B) \, = \, - {\rm i} \, V_d \, 
\frac{\lambda^2}{(4 \pi)^d} \,
\frac{(N - 1)^2}{4} \int_0^\infty d t_1 d t_2 d t_3
\, {\rm e}^{- m^2 (t_1 + t_2 + t_3)} \, \Delta_0^{- \frac{d}{2} + 1} \,
\Delta_B^{-1} \, ,
\label{twores}
\eeq
where we have defined
\beq
\Delta_B = \frac{1}{B^2} \sinh (B t_2) \sinh (B t_3) + \frac{t_1}{B}
\sinh \left[ B \left(t_2 + t_3 \right) \right]~,
\label{delb}
\eeq
while $\Delta_0 = \lim_{B \to 0} \Delta_B = t_1 t_2 + t_1 t_3 + t_2
t_3$.  Here we labelled by $t_2$ and $t_3$ the Schwinger parameters
associated with the $\bxi$ propagators. The second diagram in
Fig.~(\ref{irre}) is identical in form to the first one, with the same
symmetry factor. The only change is the color factor, so that the
resulting contribution to the effective action is identical to
\Eq{twores}, with the replacement $(N - 1)^2 \to N - 1$.

\begin{figure}
\begin{center} 
\leavevmode
\hbox{%
\epsfxsize=12cm
\psfrag{xi}{$\bxi$}
\psfrag{pi}{$\Pi$}
\psfrag{si}{$\sigma$}
\epsfbox{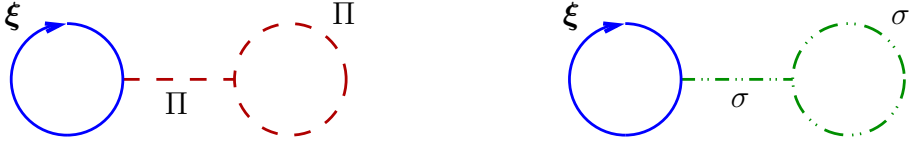}}
\end{center}
\caption{Two-loop reducible diagrams with charged propagators.}
\label{redu}
\end{figure}

Given our Feynman rules for vertices and propagators, it is of course 
straightforward to compute two-loop reducible vacuum diagrams as well.
Although these diagrams do not contribute to the effective action, it is
interesting to study them, since they can also be derived from string 
theory, as shown in \secn{low}. Specifically, we give here the results
for the two diagrams depicted in Fig.~(\ref{redu}), characterized by the
fact they have just one charged propagator.

The Feynman rule for the $\Pi^3$ vertex with our conventions is just
${\rm i} \lambda d_{a b c}$, with $d_{a b c}$ the symmetric $U(N
- 1)$ structure constants, while the $\sigma^3$ vertex is $\sqrt{2}
{\rm i} \lambda$. The calculation of the first diagram in
Fig.~(\ref{redu}) yields then
\beqa
W_{\xi \Pi}^{(2, {\rm red})} (m, B) & = & - \lambda^2 
\frac{(N - 1)^2}{2} \int d^d x \, d^d y \, G_\xi \left( x, x \right) 
\, G_\Pi \left( x, y \right) \, G_\Pi \left( y, y \right) 
\nonumber \\
& = & - \, {\rm i} \, V_d \, \frac{\lambda^2}{(4 \pi)^d}
\frac{(N - 1)^2}{2} \int_0^\infty d t_1 d t_2 d t_3
\, {\rm e}^{- m^2 (t_1 + t_2 + t_3)} \nonumber \\
& & \, \times \, \left(t_1 t_2 \right)^{- d/2} \,
\frac{B t_2}{\sinh (B t_2)}~, 
\label{redres}
\eeqa
where use was made of the fact that the symmetry factor also in this
case is $1/2$.  Again, the second diagram of Fig.~(\ref{redu}) differs
from \Eq{redres} only in the substitution $(N - 1)^2 \to N -
1$. Similar results can be derived for the two reducible diagrams with
two charged propagators.

\sect{The low-energy limit}
\label{low}

We now wish to consider the low-energy limit of the string partition
function in \Eq{effstr}, at genus $g = 1,2$, in order to isolate the
contribution of charged scalars circulating in the loops. As we shall
see, although for the bosonic string these scalars are tachyons, it is
possible to recover exactly the expressions derived with field theory
methods in \secn{field}, on a diagram by diagram basis and for
arbitrary space-time dimension $d$.

The basic idea of the field theory limit for a string amplitude or
effective action is to trade the moduli describing the shape of the
Riemann surface for dimensionful quantities, measuring the size of
various sections of the string diagram in units of $\a'$. The
logarithms of the multipliers of Schottky transformations, for
example, are associated with the length of the corresponding loops by
setting $\log (k_\mu ) = - T_\mu/\a'$, where $T_\mu$ is the sum of the
Schwinger parameters associated with the propagators forming the
$\mu^{\rm th}$ loop. It is also straightforward to identify the
contributions of states belonging to different mass levels of the
string circulating in a given loop: the operator formalism, in fact,
shows that each mass level corresponds to a given power of the
multiplier in a Taylor expansion of the integrand for small
$k_\mu$. For the bosonic string, this expansion starts with
$k_\mu^{-2}$, a sign of the tachyonic instability. This singularity
can however be readily regularized by recalling that the tachyon mass
squared is $m^2 = -1/\a'$ and setting~\cite{DiVecchia:1996kf}
\beq
\frac{d k_\mu}{k_\mu^2} = - \frac{1}{\a'} \exp \left(\frac{T_\mu}{\a'}
\right) d T_\mu = - \frac{1}{\a'} \exp \left( - m^2 T_\mu \right) d T_\mu~.
\label{tac}
\eeq
The external field is a source of further powers of $\a'$: in fact,
the field $F_{12}^{(\mu)} = \tan(\pi \epsilon_\mu)$ introduced in
\secn{backem} is dimensionless, while we want to take the low energy
limit keeping fixed the physical, dimensionful field $B$. Below, we
will always concentrate on the case in which only one boundary is
charged, setting $\e_g \equiv \e \neq 0$. The field theory limit
is then defined by $\tan(\pi \e) = 2 \pi \a' B$, which implies $\e = 2
\a' B + {\cal O} \left(\a'^3 \right)$. Finally, one must introduce in
\Eq{effstr} the appropriate overall factor, consistent with unitarity
and containing the appropriate power of the string coupling. To this
end, we follow the conventions of~\cite{DiVecchia:1996uq} and
normalize the string diagrams with an overall constant $C_g$, given
by~\eq{norms}. The string coupling $g_S$ must also be matched with the
scalar self-coupling $\lambda$, which can be done by computing a
simple tree-level amplitude from the Lagrangian \eq{lag1} and
comparing with the result obtained from string theory, as done
in~\cite{Frizzo:1999zx}. The results are
\beq C_g =
\frac{1}{(2 \pi)^{d g}} \,\, g_S^{2 g - 2} \, \left( 2 \a' \right)^{-
d/2}~, \qquad g_S = \frac{1}{4} \, \lambda \, \left( 2 \a' \right)^{(6
- d)/4}~.
\label{norms}
\eeq
To illustrate the procedure, we begin by deriving from \Eq{effstr} the
one-loop effective action, \Eq{oneres}.

\subsection{One loop}
\label{olo}

It is straightforward to write down an explicit expression for the
partition function in \Eq{effstr} evaluated at one loop. In this case,
the period matrix is just a number, simply related to the multiplier
$k$ by $2 \pi \ii \tau = \log k$. Next, observe that for $g = 1$ there
are no holomorphic Prym differentials, so that the matrix $\tau_\e$
is just the number given by the second expression in \Eq{taue},
\beq
\tau_\e = \ex{2 \pi \ii \e \tau} - 1 = k^\e - 1~.
\label{taue1}
\eeq
The ratio ${\cal R}$ defined in \Eq{ratioclosed} also simplifies
considerably, since at one loop there is only one primary class in the
Schottky group (the one represented by the single generator $S(z)$,
which can be taken to act as $S(z) = k z$ by fixing the overall
projective invariance). One finds then
\beq
{\cal R}_1 \left(k, \e \tau \right) = \prod_{n = 1}^\infty
\frac{(1 - k^n)^2}{(1 - k^{n - \e})(1 - k^{n + \e})}~.
\label{rat1}
\eeq
At one loop, in order to get the effective action, one must introduce an
additional factor of $(-\log k)^{-1}$ in the integration measure,
exactly as was done in the field theory computation. Notice that the
factor of $1/2$ present in the definition of the $1$-loop effective
action cancels against the contributions related to the two possible
orientations of the open strings. Putting together all the
ingredients we find
\beq
Z_F (1) = \ii\, C_1 \frac{\tan \left( \pi \e \right)}{\pi} 
\int\limits_0^1 \frac{d k}{k^2} \, \frac{k^{\e (1 - \e)/2}}{k^\e - 1}
\left(- \frac{\log k}{2 \pi} \right)^{-\frac{d}{2}} \prod_{n = 1}^\infty 
\frac{(1 - k^n)^{4 - d}}{(1 - k^{n - \e})(1 - k^{n + \e})}~, 
\label{part1}
\eeq
which is the results of Ref.~\cite{Bachas:1992bh}, for the magnetic
case. Recovering \Eq{oneres} is now straightforward: one must change
variables according to $k = \exp(- t/\a')$, as noted above; then,
substituting Eqs.~(\ref{tac}) and (\ref{norms}) and setting $\tan(\pi
\e) = 2 \pi \a' B$, one observes that the overall power of $\a'$
cancels, a necessary condition for the field theory limit to be well
defined. Expanding in powers of $k$ and retaining only the leading
power (all subleading powers are now exponentially suppressed as $\a'
\to 0$) one finally recovers \Eq{oneres}, with the exact normalization
factor, except for the `color' factor $N - 1$. This factor is easily
understood in terms of the $D$-brane picture of the string
calculation: as shown in Fig.~(\ref{branes}), strings contributing to
\Eq{effstr} stretch between a charged $D$-brane and $N - 1$ neutral
$D$-branes, building up the fundamental representation of $U(N)$,
broken to $U(N - 1)$ by the choice of background. At one loop, there
is only one string, and thus $N - 1$ possible attachments to the
neutral branes. \Eq{oneres} is thus reproduced exactly, including the
correct dependence on the space-time dimension $d$.

\subsection{Two loops}
\label{tlo}

The field theory limit of string amplitudes appropriate to recover
Feynman diagrams for adjoint scalars at two loops was studied in
detail in Ref.~\cite{Frizzo:1999zx,Marotta:1999re}. Here we 
briefly summarize the general features of the method, and then focus
on the application to the new quantity arising in the presence of an
external field, the matrix $\tau_\ve$.

At two loops, vacuum diagrams with cubic vertices have two possible
topologies, depicted in Fig.~(\ref{irre}) and in Fig.~(\ref{redu}).
As discussed in \cite{Frizzo:1999zx}, and having fixed projective
invariance as described in~\secn{vacen}, a change of variables
appropriate to isolate the corner of moduli space associated with the
irreducible diagram in Fig.~(\ref{irre}) is given by
\beq
k_1 = \exp \left( - \frac{t_1 + t_3}{\a'} \right)~, \quad
k_2 = \exp \left( - \frac{t_2 + t_3}{\a'} \right)~, \quad
\eta = \exp \left( - \frac{t_3}{\a'} \right)~,
\label{chirre}
\eeq
where $t_i$ are the Schwinger parameters associated with the three
propagators in the diagram. Referring to the open string diagram in
Fig.~(\ref{branes}), this choice of variables corresponds to the
assignement of proper times $t_2$ and $t_3$ to the loop with a charged
boundary, with $t_3$ associated with the propagator shared with the
other loop.  The integration region in moduli space is over all
inequivalent surfaces. In the field theory limit it is
determined~\cite{DiVecchia:1996kf} by requiring that the surface be
non-singular, and by taking into account the symmetry under the
exchange $k_1 \leftrightarrow k_2$ (part of the `residual modular
group'~\cite{Bianchi:1989du}).  In terms of the field theory
parameters introduced in~\eq{chirre} one finds simply $0 < t_3 < t_2 <
t_1 < \infty$.

The reducible diagram in Fig.~(\ref{redu}), on the other hand, arises
from the other singular corner of moduli space, which can be
parametrized by picking
\beq
k_1 = \exp \left( - \frac{t_1}{\a'} \right)~, \quad
k_2 = \exp \left( - \frac{t_2}{\a'} \right)~, \quad
\eta = 1 - \exp \left( - \frac{t_3}{\a'} \right)~,
\label{chredu}
\eeq
with the integration region, in the field theory limit, given by
$0 < t_2 < t_1< \infty$ and $0< t_3 < \infty$.

The expansion in powers of $k_\mu$ of quantities expressed in terms of
infinite series or products over the Schottky group, such as the
period matrix or the differentials $\zeta_\mu^{\vec{\e}}$ is greatly
simplified by the fact that higher-order Schottky transformations,
such as, say $S_\mu^n$, contribute to projective invariant quantities
terms of order $k_\mu^n$.  Since here we are interested in the lowest
order in the expansion, in principle we can then discard all
contributions from Schottky transformations other than the identity.
In the presence of an external field, however, this is not quite true,
since the matrix elements of $\tau_\ve$ are expressed as integrals over
the $b$-cycles of the surface, and thus they can receive leading order
contributions even from terms in the integrand arising from first
order Schottky transformations. To illustrate this fact, consider, say,
the term in the series defining $\zeta^{\vec{\e}}_2(z)$ which
involves the transformation $S_1$. One can write
\beqa
\frac{1}{z - S_1(\eta)} - \frac{1}{z - S_1(1)} & = & 
\frac{d}{d z} \log \left[ \frac{z - S_1(\eta)}{z - S_1(1)} \,
\frac{S_1(x_0) - S_1(\eta)}{S_1(x_0) - S_1(1)} \right] \nonumber \\
& = & \frac{d}{d z} \log \left[ \frac{S_1^{-1}(z) - \eta}{S_1^{-1}(z) - 1} 
\, \frac{x_0 - \eta}{x_0 - 1} \right]~,
\label{harm}
\eeqa
where we have introduced and arbitrary point $x_0$, and we made use 
of the projective invariance of harmonic ratios such as the one appearing 
as argument of the logarithm. Clearly, upon performing a definite integration
with a limit of the form $S_1(w)$ as in \Eq{taue}, \Eq{harm} can give 
a contribution of order zero in the multipliers.

Let us now focus on the main new ingredient in \Eq{effstr}, the
determinant of the twisted period matrix $\tau_\ve$. Using the
definition \eq{taue}, it can readily be written as
\beq
\det \left( \tau_\ve \right) = \frac{1}{2 \pi {\rm i}} \int_y^{S_1(y)}
\hspace{-12pt} d z ~\ex{ 2 \pi {\rm i} \vec{\e} \cdot \vec{\Delta}_z} 
\Big[ \left( {\rm e}^{2 \pi {\rm i} \left(\vec{\e} 
\cdot \tau \right)_2} - 1 \right) \zeta_1^{\vec{\e} \cdot \tau} (z) - 
\left( {\rm e}^{2 \pi {\rm i} \left(\vec{\e} \cdot \tau \right)_1} - 1 
\right) \zeta_2^{\vec{\e} \cdot \tau} (z) \Big],
\label{dete1}
\eeq
where $y$ is an arbitrary point of the surface.  For the
$\zeta_\mu^\ve$ differentials, we can use \Eq{zeta}, retaining in the
sums only the terms arising from $T_\a = {\bf 1}$ and $T_\a = S_1$, as
discussed above. The result, with generic fixed points $\eta_\mu$ and
$\xi_\mu$, is
\beqa
\zeta_1^{\vec{\e} \cdot \tau} (z) & \simeq & \frac{ {\rm e}^{2 \pi 
{\rm i} \left(\vec{\e} \cdot \tau \right)_1}}{z - \eta_1} - 
\frac{1}{z - \xi_1} - \left( 1 - {\rm e}^{2 \pi {\rm i} \left(\vec{\e} 
\cdot \tau \right)_1} \right) \frac{{\rm e}^{2 \pi {\rm i} \left(\vec{\e} 
\cdot \tau \right)_1}}{z - S_1(\eta_1)}~,
\nonumber \\
\zeta_2^{\vec{\e} \cdot \tau} (z) & \simeq & \frac{{\rm e}^{2 \pi {\rm i} 
\left(\vec{\e} \cdot \tau \right)_2}}{z - \eta_2} -
\frac{1}{z - \xi_2} - \frac{{\rm e}^{2 \pi {\rm i} \left(\vec{\e} 
\cdot \tau \right)_1}}{z - S_1(\xi_2)} + \frac{{\rm e}^{2 \pi {\rm i} 
\left(\left(\vec{\e} \cdot \tau \right)_1 + \left(\vec{\e} \cdot \tau 
\right)_2 \right)}}{z - S_1(\eta_2)}~,
\label{lowzeta}
\eeqa
where we ignored all $z_0$-dependent terms, since they cancel out in
the determinant~\eq{dete1}.  \Eq{lowzeta} can be simplified by making
use of the fact that $S_1(\eta_1) = \eta_1$, and further by
implementing our choice of fixed points, $\eta_1 = 0$, $\xi_1 =
\infty$, $\xi_2 = 1$, which implies that the explicit form of $S_1$ is
just $S_1(z) = k_1 z$. At this point we must also specify the diagram
we are considering: we begin with the irreducible contribution of
Fig.~(\ref{irre}). Inserting $\vec{\e} = (0, \e)$, and substituting
our parametrization for this diagram, \Eq{chirre}, we find
\beqa
\zeta_1^{\vec{\e} \cdot \tau} (z) & \simeq & \frac{1}{z} 
{\rm e}^{- 4 B t_3}~,
\nonumber \\
\zeta_2^{\vec{\e} \cdot \tau} (z) & \simeq & \frac{{\rm e}^{- 
2 B (t_2 + t_3)}}{z - \eta} - \frac{1}{z - 1} - \frac{{\rm e}^{- 
2 B t_3}}{z - k_1} + \frac{{\rm e}^{- 2 B (t_2 + 2 t_3)}}{z - k_1 \eta}~.
\label{finzeta}
\eeqa
The last ingredient to be expanded is the Riemann class~\eq{ricla}. At
$g = 2$, retaining again only the contributions of $T_\a = \{{\bf 1},
S_1\}$ and using \Eq{chirre} we find
\beq
\exp \left[ 2 \pi {\rm i} \vec{\e} \cdot \vec{\Delta}_z \right] \simeq 
{\rm e}^{B (t_2 + t_3)} \left[ \frac{z - 1}{z - 
\eta} \, \frac{1 - k_1 \eta}{1 - k_1} \, \frac{z - k_1}{z - k_1 \eta} 
\right]^\e \, {\rm e}^{\ii \pi \e}~. 
\label{expdel}
\eeq
The expression for the determinant of $\tau_\ve$ is then
\beqa
\det \left( \tau_\ve \right) & \simeq & \frac{{\rm e}^{ \ii \pi 
\e}}{2 \pi {\rm i}} \, {\rm e}^{B (t_2 + t_3)} \left( \frac{1 - k_1 \eta}{1 - 
k_1} \right)^\e \int_\eta^{k_1 \eta} d z \left( \frac{z - 1}{z - \eta} 
\frac{z - k_1}{z - k_1 \eta} \right)^\e \nonumber \\ & \times & \Bigg[ 
\left( {\rm e}^{- 2 B (t_2 + t_3)} - 1 \right) \frac{{\rm e}^{- 4 B t_3}}{z}
- \left( {\rm e}^{- 2 B t_3} - 1 \right) \label{dete2}  \\ && \,\, \, \times
\left( \frac{{\rm e}^{- 2 B (t_2 + t_3)}}{z - \eta} - \frac{1}{z - 1} - 
\frac{{\rm e}^{- 2 B t_3}}{z - k_1} + \frac{{\rm e}^{- 2 B (t_2 + 
2 t_3)}}{z - k_1 \eta} \right) \Bigg]~,
\nonumber
\eeqa
where we picked $y = \eta$, so that both limits of integration
coincide with branching points of the integrand. One observes that
$\det \left(\tau_\ve \right)$ is a linear combination of integrals of
the form
\beq
I(a) \equiv - \int_{k_1 \eta}^\eta \, \frac{d z}{z - a} \left( 
\frac{z - 1}{z - \eta}  \frac{z - k_1}{z - k_1 \eta} \right)^\e = {\rm
e}^{- {\rm i} \pi \e} I_1 (a) + I_2 (a)~,
\label{inta}
\eeq
where $a$ can take the values $\{0, k_1 \eta,k_1, \eta,1 \}$. Since in
the limit we are considering one has $0 < k_1 \eta \ll k_1 \ll \eta
\ll 1$, the integrand has a cut for $z < k_1$, which we have made
explicit by defining $I_1(a)$ as the integral ranging between $k_1
\eta$ and $k_1$, while $I_2 (a)$ ranges between $k_1$ an $\eta$. The
phase multiplying $I_1(a)$ depends on the choice of Riemann sheet
beyond the branch point $z = k_1$, and does not affect the final
result in the limit $\a' \to 0$.

To proceed, we need to know to what order in the $\a'$ expansions these 
integrals need to be computed. Inspection of Eqs.~\eq{effstr}, \eq{norms} and
\eq{chirre} shows that we only need the keep singular terms, which behave 
as $\a'^{-1}$. With this accuracy one finds
\beqa
I_1 (1) \simeq I_1 (\eta) \simeq 0~; && \qquad  I_1 (k_1) \simeq I_1 
(k_1 \eta) \simeq \frac{1}{2 \a' B} {\rm e}^{2 B t_3}~; \nonumber \\
I_1 (0) & \simeq & {\rm e}^{2 B t_3} \, \frac{1 - {\rm e}^{2 B 
t_3}}{2 \a' B}~,
\label{i1}
\eeqa
for the various $I_1$ integrals, while the results for $I_2$ are
\beqa
I_2 (k_1 \eta) \simeq I_2 (0) \simeq - \frac{{\rm e}^{2 B t_3}}{\a'} t_1~; 
&& \qquad  I_2 (\eta) \simeq - \frac{1}{2 \a' B} {\rm e}^{2 B t_3}~; 
\nonumber \\ I_2 (1) \simeq 0~; && \qquad
I_2 (k_1) \simeq  - \frac{{\rm e}^{2 B t_3}}{\a'} \left( \frac{1}{2 B}
+ t_1 \right)~.
\label{i2}
\eeqa
Substituting these results into \Eq{dete2}, and keeping only the
leading power in $\a'$, we find finally
\beq
\det \left( \tau_\ve \right) \, = \, \frac{1}{{\rm i} \pi \a'} \left[ 
t_1 \sinh \left[ B (t_2 + t_3) \right] + \frac{\sinh \left( B t_2 
\right) \sinh \left( B t_3 \right)}{B} \right] \, + \, {\cal O} 
\left((\a')^0 \right)~,
\label{dete3}
\eeq
precisely the same structure arising in field theory, as seen by
comparing to \Eq{twores}. With a straightforward computation of the
measure $d Z_2$ in this limit, noting that the ratio ${\cal R}_2$
contributes just a factor of unity, and assembling all normalization
factors, the string-derived result for the effective action is
\beqa
W_{{\rm st}}^{(2)} (m, B) & = & V_d \, \frac{\lambda^2}{(4 \pi)^d} \,
\frac{(N - 1)^2}{2} \int_0^\infty d t_1 \int_0^{t_1} d t_2 \int_0^{t_2} 
d t_3 \nonumber \\
&& \, \times \, {\rm e}^{- m^2 (t_1 + t_2 + t_3)} \, \Delta_0^{- d/2 + 1} 
\, \Delta_B^{-1}~.
\label{stres}
\eeqa
This result maps exactly onto \eq{twores}. The color factor $(N -
1)^2$ arises in string theory because at two loops we have two strings
attached at one end to the charged brane, while the two free ends can
choose between $N - 1$ neutral branes. Further, to get the complete
result we must add to \eq{stres} the two contributions corresponding
to charging the other two boundaries of the double annulus: they
symmetrize the integrand, so that one can complete the integration
region including a factor of $1/2$, reproducing the exact
normalization. Finally, the missing factor of $\ii$ in \eq{stres} is
due to the fact that string amplitudes are normalized to give directly
the $T$-matrix element~\cite{DiVecchia:1996uq}, while Feynman diagrams
give the $S$-matrix element. Once again the result is correct for
arbitrary $d$\footnote{For a justification of the fact that the $d$
  dependence is correct even if $d \neq 26$,
  see~\cite{Frizzo:2000ez}}. Notice that the string calculation
described in \secn{backem} corresponds only to the first diagram in
Fig.~(\ref{irre}): to get a singlet field $\sigma$ propagating in the
string picture we would need a string with both ends on the charged
brane, a configuration corresponding to having {\it two} charged
boundaries with the same value of $\e$.

To conclude, we briefly show that one can also recover exactly the
reducible diagrams in Fig.~(\ref{redu}). Again, the string computation
in the chosen configuration will yield only the diagram involving the
propagation of $\Pi$ fields (the first in Fig.~(\ref{redu})), but a
very similar calculation would reproduce the $\sigma$ diagram as well.

The calculation in this case is simplified by the fact that the
ordinary period matrix $\tau$ becomes diagonal in the limit defined by
\Eq{chredu}, as $\a'\to 0$. As a consequence, the off-diagonal matrix
element $\left( \tau_\ve\right)_{2 1}$ also vanishes, and thus $\det
\tau_\ve$ involves only $\zeta^{\vec{\e}}_1$, which is simpler in the
field theory limit, as shown by \Eq{finzeta}. A short calculation yields
\beq
\det \left( \tau_\ve \right) \, = \, \frac{1}{{\rm i} \pi} \, \sinh 
\left( B t_2 \right) \, \frac{t_1}{\a'} \, + \, {\cal O} 
\left((\a')^0 \right)~,
\label{detered}
\eeq
Inserting all normalization factors we find then
\beqa
W_{{\rm st}}^{(2, {\rm red})} (m, B) & = & V_d \, 
\frac{\lambda^2}{(4 \pi)^d} \frac{(N - 1)^2}{2} \int_0^\infty d t_1 
\int_0^{t_1} d t_2 \int_0^\infty d t_3
\, {\rm e}^{- m^2 (t_1 + t_2 + t_3)} \nonumber \\
& \times & \left(t_1 t_2 \right)^{- d/2 + 1} \,
\frac{B}{t_1 \sinh (B t_2)}~, 
\label{redstres}
\eeqa 
which has the right structure to reproduce \Eq{redres}.  It is
interesting to observe how the combinatoric factors arising from the
string match those computed in field theory. Here again we could let
any one of the three boundaries of the double annulus be the charged
one, however when the charged boundary is the external one the
corresponding field theory diagram is different: the charged fields
$\bxi$ are now propagating in both loops, so that the propagator
joining them must be a $\sigma$ propagator.  If we want to recover the
first diagram in Fig.~(\ref{redu}), we must add to \Eq{redstres} only
the configuration with $t_1 \leftrightarrow t_2$; this symmetrizes the
integrand, so that the integration region can be extended to match
\Eq{redres}, with precisely the required overall factor.  This
completes our proof that {\it all} two-loop vacuum diagrams for
adjoint scalars in a constant background field can be precisely
recovered from bosonic string theory: the remaining diagrams,
involving the coupling of $\bxi$ with $\sigma$, can be similarly
reproduced starting with the appropriate modification of \Eq{effstr}.

\sect{Conclusions}
\label{conclusions}

Bosonic open strings in the presence of a constant gauge field
strength provide an interesting system to study. On the technical side
they display the general features of more complicated models, in
particular the presence of cuts along some cycles of the Riemann
surface representing the multiloop string amplitude. One can then use
this simpler system to study objects like twisted determinants and
differentials that will appear also in various other contexts. On the
other hand, charged open strings are physically a very interesting
system, primarily because they are directly related to perturbative
gauge theories. As an example of this relation, we have shown that the
Euler-Heisenberg effective action for a gauge field is naturally
encoded in the string result.

A direct generalization of the computation presented in this paper
would be to derive the Euler-Heisenberg effective action for pure
Yang-Mills theory. In order to do this, there is no need to modify the
string construction and the starting point is always~\Eq{effstr}. What
needs to be changed is the definition of the field theory limit, in
order to isolate the contributions to the loop integrals of the first
excited state in the spectrum of the open bosonic string, which is a
massless vector. In practice, this means that the expansion in
multipliers of the various geometrical objects appearing in
\Eq{effstr} has to be pushed one order higher. Hopefully, this
computation will provide a simpler setup where it might be possible to
see explicitly Yang-Mills Feynman diagrams arising from the field
theory limit of string amplitudes beyond one loop. Another interesting
aspect is to study the relation between our computations and the the
world-line formalism (see~\cite{Schubert:2001he} for a recent detailed
review and references). The world-line formalism has already been
applied succesfully to the study of Euler-Heisenberg effective action
at one loop and beyond, see for
instance~\cite{Reuter:1996zm,Kors:1998ew,Sato:2000cr}.

At the string level, it would be very interesting to extend our
computation to the case of superstrings and complete the multiloop
extension of the result of~\cite{Bachas:1992bh}. On the other
hand, the supersymmetric setup can be T-dualized to get the multi-body
interaction among D-branes in type II theories. There are many results
available on this problem derived by using different approaches, like
11D supergravity and M(atrix) theory. It would clearly be interesting
to see whether an exact string computation can bring some further
insight into this problem.

Let us conclude with a general comment on the two different
formulations we gave for the charged open string partition
function,~\Eq{clch} and~\Eq{effstr}. Even if the two formulae are
exactly equivalent at the string level, the gauge theory results can
be derived only by performing the field theory limit of the second
equation. The reason is clear: the expansion of~\Eq{clch} in the
multipliers $q$ organizes the result by separating the contributions
to the string diagram coming from the exchange of {\em closed string}
states between D-branes; in the open string channel, on the other
hand, one uses~\Eq{effstr}, where different powers of $k$ are related
to the propagation of {\em open string} states. Thus the first
formulation is useful if one wants to study at low energies a
gravitational theory, while the second formulation is the one we used
to derive the gauge theory results. In general, the modular
transformation that maps the two equations into each other mixes all
terms of the $q$ and $k$ expansions, so that the two low-energy
results are completely unrelated. In special cases, however, it may
happen that the series in $q$ reduces to a single term, which then has
to match what is found in the $k$ expansion of the open string
formula. In these cases, then, the gauge theory result can be directly
derived from the closed string (or ``gravitational'') description
(see~\cite{DiVecchia:2003ae} for an explicit discussion of some such
examples).  It would be very interesting to see whether a similar
situation can occur also at two loops, though clearly this is not the
case for bosonic strings. If an example of this type were to exist
also for $g = 2$, it could help to gain a better understanding of the
perturbative aspects of the gauge/gravity correspondence.

\vspace{1cm}

\noindent {\large {\bf Acknowledgments}}

\vspace{3mm}
\noindent This work is partially supported by the European Commission,
under RTN program MRTN-CT-2004-0051004 and by the Italian MIUR under
the contract PRIN 2003023852.  We would like to thank C.-S. Chu for
collaboration during the early stages of this project, as well as
L. Alvarez-Gaum\'e and P. Di Vecchia for interesting discussions and
suggestions. S.~Sc.  thanks the CERN PH Department (TH Division) for
hospitality during part of this work.

\vspace{2cm}

\appendix

\section{Relating different representations of the two-loop partition 
function}
\label{appa}

For completeness, we give here a brief discussion of the derivation of
\eq{scg} and \eq{bm} from \eq{fp}.  The steps connecting~\eq{fp}
and~\eq{scg} are carefully explained by Roland
in~\cite{Roland:1993pm}. The quadratic differentials used are those
derived in~\cite{DiVecchia:1989id} and their overlap with the Beltrami
differentials related to the moduli $k_i$ and $\eta$ is explicitly
computed in the Schottky parametrization. Actually this derivation is
valid beyond two loops and Roland obtains from~\eq{fp} the general
expression of~\cite{DiVecchia:1987uf}. As a remark, notice that the
formulae in~\cite{Roland:1993pm} do not contain explicitly the
normalization factor $\sqrt{{\rm det}\langle\phi_j|\phi_k
\rangle}$. This is simply because the sewing procedure selects a
particular form for the quadratic differentials and the resulting
determinants, written in the Schottky parametrization
\cite{DiVecchia:1989id,Russo:2003tt}, already include this
normalization.

The connection between~\eq{fp} and~\eq{bm}, on the other hand, has
been explicitly studied in Section~7.1 of~\cite{D'Hoker:2001qp}. The
basic idea is to choose as basis for the $\phi$'s simply the set of
products of the usual Abelian differentials $\omega_\mu$. We will denote
this particular basis as $\Omega_1(z) = (\omega_1(z))^2$, $\Omega_2(z)
= (\omega_2(z))^2$ and $\Omega_3(z) = \omega_1(z) \omega_2(z)$. The
important property of this choice is that it naturally parametrizes
the variations of the period matrix under changes of the
$2$-dimensional metric $g$ (see for
instance~\cite{Verlinde:1986kw}). In fact, for example, $\delta
\tau_{i j}/\delta g^{zz} \sim \omega_i(z) \omega_j(z)$. From this
fact, we can see that $\delta \tau_{i j}/\delta m^a = \langle \mu^a |
\delta \tau_{i j}/\delta g^{z z} \rangle \sim \langle \mu^a |
\omega_i(z) \omega_j(z) \rangle $, where $\mu^a$ is the Beltrami
differential associated to the modulus $m^a$. By identifying the
elements of the period matrix with the moduli $m^a$, we immediately
see that the overlap $\langle \mu^{\tau_{k l}} | \Omega_I(z) \rangle $
is simply proportional to the identity matrix. Then, by using the
bosonization equivalence on a surface of genus $g = 2$, one can
reexpress the fermionic determinant in terms of $\theta$-functions and
other geometrical objects, as
\beq
\frac{ {\rm det}' (\partial^\dagger \partial) }{ \sqrt{\det
\langle \Omega_j | \Omega_k \rangle} } =
\frac{\prod_{I < J} E(z_I, z_J) \prod_{I = 1}^{3} \sigma (z_I)^3} 
{Z_1 \, \det \Big( \Omega_I (z_J) \Big)}
\; \theta \! \left( 3 \vec{\Delta} - \sum_{I = 1}^{3} \vec{J}(z_I)
\, \Bigg| \, \tau \right)\;,
\label{bosQ3}
\eeq
where the prime form $E(z,w)$, the function $\sigma(z)$, the Riemann
class $\vec{\Delta}$ and the Jacobi map $\vec{J}$ are defined as in
Ref.~\cite{Russo:2003tt}.  Note that although the right hand side of
\Eq{bosQ3} appears to depend on $g + 1 = 3$ coordinates $z_I$ on the
Riemann surface, the equality with the left hand side shows that this
dependence actually cancels out in the ratio.

By using~\eq{bosQ3} inside~\eq{fp} one obtains Eq.~(7.2)
of~\cite{D'Hoker:2001qp} and then the expression~\eq{bm} originally
proposed in~\cite{Belavin:1986tv,Moore:1986rh}. This proves the
equivalence between~\eq{bm} and~\eq{scg}.

\section{Scalar propagator in a background field}
\label{appb}

In our chosen gauge, $A_\mu = B x_1 g_{\mu 2}$, the propagator for the
charged field $\bxi$ in $d$ space-time dimensions must satisfy the
equation
\beq
\left( \Box + m^2 + 2 {\rm i} B x_1 \partial_2 + B^2 x_1^2 \right) 
G_\xi (x,y) = - {\rm i} \, \delta^d (x - y)~.
\label{kineq}
\eeq
This equation can be easily solved by Fourier transforming with
respect to all coordinates $x_\mu$ except $x_1$. To keep our notation
simple, we will denote Fourier transforms with the same symbol as the
original function. The partial Fourier transform of $G_\xi (x,y)$
obeys the equation
\beq
\left[ \partial_1^2 + k_0^2 - \left|\bkp \right|^2 - m^2
- \left( B x_1 + k_2 \right)^2 \right] G_\xi (k_0,\bkp; x_1, y_1)
= {\rm i} \, \delta^d (x_1 - y_1)~.
\label{foueq}
\eeq
One recognizes in the $x_1$-dependent terms of the kinetic operator the 
quantum mechanical hamiltonian of a harmonic oscillator with mass $M = 1/2$
and angular frequency $\Omega = 2 |B|$, oscillating around the point
$x_1 = - k_2/B$. Exploiting the completeness of the eigenfunctions of this 
hamiltonian one is immediately lead to the representation
\beqa
G_\xi (k_0,\bkp; x_1, y_1) & = & \sum_{n = 0}^\infty \psi_n^* \left( y_1 +
\frac{k_2}{B} \right) \psi_n \left( x_1 + \frac{k_2}{B} \right)
\nonumber \\
& \times & \frac{{\rm i}}{k_0^2 - \bkp^2 - m^2 - \left(2 n + 1\right) |B|}~,
\label{hoc}
\eeqa
where
\beq
\psi_n (z) = \sqrt{\frac{1}{2^n n!} \sqrt{\frac{|B|}{\pi}}} \,
\, H_n \left(\sqrt{|B|} \, z \right) \, \exp \left(- |B| z^2/2 \right)
\label{hops}
\eeq
are the appropriate eigenfunctions, with $H_n$ the Hermite polynomials.
It is now straightforward to Fourier transform also with respect to the 
remaining coordinates $x_1$ ad $y_1$, recalling that the momentum space
eigenfunctions of the harmonic oscillator are again given by Hermite 
polynomials. Taking into account the fact that momentum is not conserved
in the $x_1$ direction, define then the momentum space propagator as
\beq
G_\xi (k_0,\bkp; k_1, k_1') = \int d x_1 d y_1
{\rm e}^{- {\rm i} (k_1 x_1 - k_1' y_1)} G_\xi (k_0,\bkp; x_1, y_1)~,
\label{momprop}
\eeq
to find explicitly
\beqa
& & \hspace{-5mm} G_\xi (k_0,\bkp; k_1, k_1') \, = \, \frac{1}{\sqrt{\pi |B|}} 
\exp \left[ - {\rm i} \, \frac{k_2}{B} (k_1 - k_1') \right] \exp 
\left[- \frac{k_1^2 + k_1'^2}{2 |B|} \right] \label{expl} \\ 
& & \hspace{5mm} \times \, \sum_{n = 0}^\infty \frac{1}{2^n \, n!} 
H_n \left(\frac{k_1}{\sqrt{|B|}} \right)
H_n \left(\frac{k_1'}{\sqrt{|B|}} \right) 
\frac{2 \pi {\rm i}}{k_0^2 - \bkp^2 - m^2 - \left(2 n + 1\right) |B|}.
\nonumber
\eeqa
Eq.~(\ref{expl}) can be written in a more elegant form, closely related 
to the results obtained with string methods, by introducing a Schwinger 
representation for the denominator, writing
\beq
\frac{1}{k_0^2 - \bkp^2 - m^2 - \left(2 n + 1\right) |B|} = - \int_0^\infty
d t \, {\rm e}^{- t \, \left( m^2 + \left(2 n + 1\right) |B| -
k_0^2 + \bkp^2 \right)}~,
\label{sch}
\eeq
and by making using of the following sum rule for Hermite polynomials, 
known as Mehler's formula~\cite{BAT},
\beq
\sum_{n = 0}^\infty \frac{H_n (x) \, H_n (y)}{n!} \left( \frac{w}{2} \right)^n
= \frac{1}{\sqrt{1 - w^2}} \, \exp \left[ - \frac{(x - y)^2 \, w}{1 - w^2} +
\frac{(x^2 + y^2) \, w}{1 + w} \right]~.
\label{meh}
\eeq
Defining $k_\pm = k_1 \pm k_1'$, one can then write
\beqa
& & \hspace{-8mm} G_\xi (k_0,\bkp; k_1, k_1') \, = \, - \, \frac{2 \pi 
{\rm i}}{\sqrt{\pi}} \, \exp \left( {\rm i} \, \frac{k_2 k_-}{B} \right) 
\int_0^\infty \exp \left[ - t \left( m^2 - k_0^2 + \bkp^2 \right) \right]
\nonumber \\ & & \hspace{3mm} \times \, \frac{1}{\sqrt{2 B \sinh 
\left( 2 B t \right)}} \, \exp \left[ - \frac{1}{4 B} \left( k_+^2 \tanh 
\left( B t \right) + k_-^2 \coth \left( B t \right) \right) \right] \, .
\label{finmom}
\eeqa
One can verify that Eq.~(\ref{finmom}) has the correct limit as $B \to 0$,
\beq
\lim_{B \to 0} G_\xi (k_0,\bkp; k_1, k_1') \, = \, (2 \pi) \delta (k_1 - k_1') 
\, \frac{{\rm i}}{k^2 - m^2} \, .
\label{limg}
\eeq
In order to compute effective actions, and to gain a better
understanding of the gauge dependence of the propagator, it is useful
to Fourier transform Eq.~(\ref{finmom}) back to coordinate space. This
last calculation is straightforward, and leads to \Eq{fincoo}.  Notice
that the phase factor in the first line of Eq.~(\ref{fincoo}) is the
expected gauge link between point $x$ and point $y$. In fact with our
choice of gauge field
\beq
\exp \left[ - \frac{{\rm i}}{2} \, B \, (x_1 + y_1) (x_2 - y_2) \right] = 
\exp \left[ - {\rm i} \int_x^y A_\mu (z) \, d z^\mu \right]\,,
\label{gau}
\eeq
where the integral is performed along the straight line joining points
$x$ and $y$.  The terms in Eq.~(\ref{gau}) which depend on both $x$
and $y$ are gauge-invariant and can be assembled into a factor of the
form $\exp \left( - {\rm i} \, {\cal F}_{\mu \nu} x^\mu y^\nu/2
\right)$. Factors depending only on $x$ or on $y$, on the other hand,
can be removed with an appropriate gauge transformation. For example,
choosing as gauge function $f(x_\mu) \equiv - x_1 x_2 B/2$ one moves
to the ``symmetric'' gauge,
\beq
A_1'(x) = - \frac{x_2}{2} B \,\,; \qquad A_2'(x) = \frac{x_1}{2} B~,
\label{sgau}
\eeq
where only the gauge-invariant contribution to Eq.~(\ref{gau})
survives.  In this gauge, in coordinate space, one can make a direct
comparison with the results of
Refs.~\cite{Ritus:1975cf,Ritus:1977iu,Ritus:1998jm}, recently
discussed in \cite{Dunne:2004nc}, finding complete agreement.


\begin{thebibliography}{10}

\bibitem{Heisenberg:1935qt}
W.~Heisenberg and H.~Euler, 
 {\em Z. Phys.} {\bf 98} (1936) 714--732.

\bibitem{Schwinger:1951nm}
J.~S. Schwinger, 
 {\em Phys. Rev.} {\bf 82} (1951) 664--679.

\bibitem{Dunne:2004nc}
G.~V. Dunne, 
{\it Heisenberg-Euler effective lagrangians: basics and extensions},  
 \href{http://xxx.lanl.gov/abs/hep-th/0406216}{{\tt hep-th/0406216}}.

\bibitem{Bachas:1992bh}
C.~Bachas and M.~Porrati, 
 {\em Phys. Lett.} {\bf B296} (1992) 77--84,
 [\href{http://xxx.lanl.gov/abs/hep-th/9209032}{{\tt hep-th/9209032}}].

\bibitem{Bachas:1995kx}
C.~Bachas, 
 {\em Phys. Lett.} {\bf B374} (1996) 37--42,
 [\href{http://xxx.lanl.gov/abs/hep-th/9511043}{{\tt hep-th/9511043}}].

\bibitem{Tseytlin:1998kw}
A.~A.~Tseytlin,
{\em Nucl. Phys.} {\bf B524} (1998) 41, 
[\href{http://xxx.lanl.gov/abs/hep-th/9802133}{{\tt hep-th/9802133}}].

\bibitem{Russo:2003tt}
R.~Russo and S.~Sciuto, 
 {\em Nucl. Phys.} {\bf B669} (2003) 207--232,
 [\href{http://xxx.lanl.gov/abs/hep-th/0306129}{{\tt hep-th/0306129}}].

\bibitem{Russo:2003yk}
R.~Russo and S.~Sciuto, 
 {\em Fortsch. Phys.} {\bf 52} (2004) 678--683,
 [\href{http://xxx.lanl.gov/abs/hep-th/0312205}{{\tt hep-th/0312205}}].

\bibitem{Verlinde:1986kw}
E.~Verlinde and H.~Verlinde, 
 {\em Nucl. Phys.} {\bf B288} (1987) 357.

\bibitem{Alvarez-Gaume:1987vm}
L.~Alvarez-Gaume, J.~B. Bost, G.~W. Moore, P.~Nelson, and C.~Vafa, 
 {\em Commun. Math. Phys.}  {\bf 112} (1987) 503.

\bibitem{Aoki:2003sy}
K.~Aoki, E.~D'Hoker, and D.~H. Phong, 
 {\em Nucl. Phys.} {\bf B688} (2004) 3--69,
 [\href{http://xxx.lanl.gov/abs/hep-th/0312181}{{\tt hep-th/0312181}}].

\bibitem{Antoniadis:2004qn}
I.~Antoniadis and T.~R. Taylor, 
 {\em Nucl. Phys.} {\bf B695} (2004) 103--131,
 [\href{http://xxx.lanl.gov/abs/hep-th/0403293}{{\tt hep-th/0403293}}].

\bibitem{Scherk:1971xy}
J.~Scherk, 
 {\em Nucl. Phys.} {\bf B31} (1971) 222--234.

\bibitem{Metsaev:1987ju}
R.~R.~Metsaev and A.~A.~Tseytlin,
{\em Nucl.\ Phys.} {\bf B298} (1988) 109.

\bibitem{Mangano:1987xk}
M.~L. Mangano, S.~J. Parke, and Z.~Xu, 
 {\em Nucl. Phys.} {\bf B298} (1988) 653.

\bibitem{Bern:1991aq}
Z.~Bern and D.~A. Kosower, 
 {\em Nucl. Phys.} {\bf B379} (1992) 451--561.

\bibitem{Bern:1993mq}
Z.~Bern, L.~J. Dixon, and D.~A. Kosower, 
 {\em Phys. Rev. Lett.} {\bf 70} (1993) 2677--2680,
 [\href{http://xxx.lanl.gov/abs/hep-ph/9302280}{{\tt hep-ph/9302280}}].

\bibitem{Bern:1996je}
Z.~Bern, L.~J. Dixon, and D.~A. Kosower, 
 {\em Ann. Rev. Nucl. Part. Sci.} {\bf 46} (1996) 109--148,
 [\href{http://xxx.lanl.gov/abs/hep-ph/9602280}{{\tt hep-ph/9602280}}].

\bibitem{DiVecchia:1996uq}
P.~Di~Vecchia, L.~Magnea, A.~Lerda, R.~Russo, and R.~Marotta, 
 {\em Nucl. Phys.} {\bf B469} (1996) 235--286,
 [\href{http://xxx.lanl.gov/abs/hep-th/9601143}{{\tt hep-th/9601143}}].

\bibitem{Frizzo:2000ez}
A.~Frizzo, L.~Magnea, and R.~Russo, 
 {\em Nucl. Phys.} {\bf B604} (2001) 92--120, 
 [\href{http://xxx.lanl.gov/abs/hep-ph/0012129}{{\tt hep-ph/0012129}}].

\bibitem{Bern:1993wt}
Z.~Bern, D.~C. Dunbar, and T.~Shimada, 
 {\em Phys. Lett.} {\bf B312} (1993) 277--284,
 [\href{http://xxx.lanl.gov/abs/hep-th/9307001}{{\tt hep-th/9307001}}].

\bibitem{Dunbar:1994bn}
D.~C. Dunbar and P.~S. Norridge, 
 {\em Nucl. Phys.} {\bf B433} (1995) 181--208, 
 [\href{http://xxx.lanl.gov/abs/hep-th/9408014}{{\tt hep-th/9408014}}].

\bibitem{Magnea:1997kv}
L.~Magnea and R.~Russo, 
{\it Two-loop gluon diagrams from string theory}
 \href{http://xxx.lanl.gov/abs/hep-th/9706396}{{\tt hep-th/9706396}}.

\bibitem{Kors:2000bb}
B.~Kors and M.~G. Schmidt, 
{\it Two-loop Feynman diagrams in Yang-Mills theory from bosonic string 
amplitudes},
 \href{http://xxx.lanl.gov/abs/hep-th/0003171}{{\tt hep-th/0003171}}.

\bibitem{DiVecchia:1996kf}
P.~Di~Vecchia, L.~Magnea, A.~Lerda, R.~Marotta, and R.~Russo, 
 {\em Phys. Lett.} {\bf B388} (1996) 65--76, 
 [\href{http://xxx.lanl.gov/abs/hep-th/9607141}{{\tt hep-th/9607141}}].

\bibitem{Frizzo:1999zx}
A.~Frizzo, L.~Magnea, and R.~Russo, 
 {\em Nucl. Phys.} {\bf B579} (2000) 379--410,
 [\href{http://xxx.lanl.gov/abs/hep-th/9912183}{{\tt hep-th/9912183}}].

\bibitem{Marotta:1999re}
R.~Marotta and F.~Pezzella, 
 {\em Phys. Rev.} {\bf D61} (2000) 106006,
 [\href{http://xxx.lanl.gov/abs/hep-th/9912158}{{\tt hep-th/9912158}}].

\bibitem{Belavin:1986tv}
A.~A. Belavin, V.~Knizhnik, A.~Morozov, and A.~Perelomov, 
 {\em JETP Lett.} {\bf 43} (1986) 411.

\bibitem{Moore:1986rh}
G.~W. Moore, 
 {\em Phys. Lett.} {\bf B176} (1986) 369.

\bibitem{DiVecchia:1987uf}
P.~Di~Vecchia, M.~Frau, A.~Lerda, and S.~Sciuto, 
 {\em Phys. Lett.} {\bf B199} (1987) 49.

\bibitem{Blau:1987pn}
S.~K. Blau, M.~Clements, S.~Della~Pietra, S.~Carlip, and V.~Della~Pietra, 
 {\em Nucl. Phys.} {\bf B301} (1988) 285.

\bibitem{Bianchi:1988fr}
M.~Bianchi and A.~Sagnotti, 
 {\em Phys. Lett.} {\bf B211} (1988) 407.

\bibitem{Petersen:1988ux}
J.~L. Petersen, K.~O. Roland, and J.~R. Sidenius, 
 {\em Phys. Lett.} {\bf B205} (1988) 262.

\bibitem{Roland:1988hg}
K.~Roland, 
 {\em Nucl. Phys.} {\bf B313} (1989) 432.

\bibitem{D'Hoker:1988ta}
E.~D'Hoker and D.~H. Phong, 
 {\em Rev. Mod. Phys.} {\bf 60} (1988) 917.

\bibitem{D'Hoker:2001qp}
E.~D'Hoker and D.~H. Phong, 
 {\em Nucl. Phys.} {\bf B639} (2002) 129--181,
 [\href{http://xxx.lanl.gov/abs/hep-th/0111040}{{\tt hep-th/0111040}}].

\bibitem{DiVecchia:1999rh}
P.~Di~Vecchia and A.~Liccardo, 
 {\em NATO Adv. Study Inst. Ser. C. Math. Phys. Sci.} {\bf 556} (2000) 1--59,
 [\href{http://xxx.lanl.gov/abs/hep-th/9912161}{{\tt hep-th/9912161}}].

\bibitem{DiVecchia:1999fx}
P.~Di~Vecchia and A.~Liccardo, 
{\it D-branes in string theory, II},
 \href{http://xxx.lanl.gov/abs/hep-th/9912275}{{\tt hep-th/9912275}}.

\bibitem{Tseytlin:1986zz}
A.~A.~Tseytlin,
{\em Phys.\ Lett.} {\bf B176} (1986) 92.

\bibitem{Tseytlin:1986ti}
A.~A.~Tseytlin,
{\em Nucl.\ Phys.} {\bf B276} (1986) 391
[Erratum-ibid. {\bf B291} (1987) 876].

\bibitem{Frau:1997mq}
M.~Frau, I.~Pesando, S.~Sciuto, A.~Lerda, and R.~Russo, 
 {\em Phys. Lett.} {\bf B400} (1997) 52--62, 
 [\href{http://xxx.lanl.gov/abs/hep-th/9702037}{{\tt hep-th/9702037}}].

\bibitem{Chu:2002nd}
C.-S. Chu, R.~Russo, and S.~Sciuto, 
 {\em Fortsch. Phys.} {\bf 50} (2002) 871--877, 
 [\href{http://xxx.lanl.gov/abs/hep-th/0201118}{{\tt hep-th/0201118}}].

\bibitem{Pezzella:1988jr}
F.~Pezzella, 
 {\em Phys. Lett.} {\bf B220} (1989) 544.

\bibitem{Losev:1989fe}
A.~Losev, 
 {\em JETP Lett.} {\bf 49} (1989) 424--426.

\bibitem{DiVecchia:1989ht}
P.~Di~Vecchia, 
{\it The sewing technique and correlation functions on arbitrary
Riemann surfaces}, 
 NORDITA-89/30-P.

\bibitem{Fock:1937dy}
V.~Fock, 
 {\em Phys. Z. Sowjetunion} {\bf 12} (1937) 404--425.

\bibitem{Weisskopf:1996bu}
V.~Weisskopf, 
{\it The electrodynamics of the vacuum based on the quantum
theory of the electron},  
in {\em Early quantum electrodynamics} (A.~Miller, ed.), Cambridge
University Press (1994), pp.~206--226.

\bibitem{Bianchi:1989du}
M.~Bianchi and A.~Sagnotti, 
 {\em Phys. Lett.} {\bf B231} (1989) 389.

\bibitem{Schubert:2001he}
C.~Schubert, 
 {\em Phys. Rept.} {\bf 355} (2001) 73--234,
 [\href{http://xxx.lanl.gov/abs/hep-th/0101036}{{\tt hep-th/0101036}}].

\bibitem{Reuter:1996zm}
M.~Reuter, M.~G. Schmidt, and C.~Schubert, 
 {\em Annals Phys.} {\bf 259} (1997) 313--365,
 [\href{http://xxx.lanl.gov/abs/hep-th/9610191}{{\tt hep-th/9610191}}].

\bibitem{Kors:1998ew}
B.~Kors and M.~G. Schmidt, 
 {\em Eur. Phys. J.} {\bf C6} (1999) 175--182,
 [\href{http://xxx.lanl.gov/abs/hep-th/9803144}{{\tt hep-th/9803144}}].

\bibitem{Sato:2000cr}
H.-T. Sato, M.~G. Schmidt, and C.~Zahlten, 
 {\em Nucl. Phys.} {\bf B579} (2000) 492--524,
 [\href{http://xxx.lanl.gov/abs/hep-th/0003070}{{\tt hep-th/0003070}}].

\bibitem{DiVecchia:2003ae}
P.~Di~Vecchia, A.~Liccardo, R.~Marotta, and F.~Pezzella, 
 {\em JHEP} {\bf 06} (2003) 007, 
 [\href{http://xxx.lanl.gov/abs/hep-th/0305061}{{\tt hep-th/0305061}}].

\bibitem{Roland:1993pm}
K.~Roland, 
 {\em Phys. Lett.} {\bf B312} (1993) 441--450.

\bibitem{DiVecchia:1989id}
P.~Di~Vecchia, F.~Pezzella, M.~Frau, K.~Hornfeck, A.~Lerda, and S.~Sciuto, 
 {\em Nucl. Phys.} {\bf B333} (1990) 635.

\bibitem{BAT}
A.~Erd{\'e}lyi and {\it et al.}, 
{\em Higher trascendental functions}.
 \newblock McGraw--Hill, New York, 1953.

\bibitem{Ritus:1975cf}
V.~I. Ritus, 
 {\em Sov. Phys. JETP} {\bf 42} (1975) 774.

\bibitem{Ritus:1977iu}
V.~I. Ritus, 
 {\em Zh. Eksp. Teor. Fiz.} {\bf 73} (1977) 807--821.

\bibitem{Ritus:1998jm}
V.~I. Ritus, 
{\it Effective lagrange function of intense electromagnetic field in QED},  
 \href{http://xxx.lanl.gov/abs/hep-th/9812124}{{\tt hep-th/9812124}}.

\end{thebibliography}

\providecommand{\href}[2]{#2}\begingroup\raggedright\endgroup

\end{document}